\documentclass[twocolumn,showpacs,preprintnumbers,amsmath,amssymb]{revtex4-1}
\usepackage{graphicx}
\usepackage{textcomp}

\begin{document}
\title{Controlling the path of discretized light in waveguide lattices}
  \normalsize
\author{Stefano Longhi 
}
\address{Dipartimento di Fisica, Politecnico di Milano, Piazza L. da Vinci
32, I-20133 Milano, Italy}

%
\bigskip
\begin{abstract}
\noindent A general method for flexible control of the path of
discretized light beams in homogeneous waveguide lattices, based on
longitudinal modulation of the coupling constant, is theoretically
proposed. As compared to beam steering and refraction achievable in
graded-index waveguide arrays, the proposed approach enables to
synthesize rather arbitrary target paths.
\end{abstract}



\maketitle

Light propagation in waveguide lattices has received a great and
continuous interest over the past few years
\cite{Christodoulides03,rev1,rev2,rev3}, with the observation of a
host of new phenomena such as optical Bloch oscillations
\cite{Christodoulides03}, Zener tunneling \cite{Trompeter06},
diffraction management \cite{man}, dynamic localization
\cite{Longhi06,Szameit09}, Rabi oscillations \cite{Rabi} , Talbot
imaging \cite{Talbot}, and Anderson localization \cite{and}, just to
mention a few. Lattice engineering enables to mold in a rather
flexible way the flow of discretized light, hence providing
altogether new opportunities for applications
\cite{Szameit09,Moison09,Belabas09,Fan09}. In spite of the
discretized behavior imposed by the lattice, light transport in {\em
homogeneous} waveguide lattices shear some common features with
optical beam propagation in homogeneous media (hereafter referred to
as continuous beam propagation). For instance, the path followed by
a discretized optical beam in a homogeneous waveguide lattice is
straight like in a homogeneous medium, and beam spreading
(diffraction) for both discretized and paraxial continuous beams is
governed by the same algebraic law \cite{Longhi09}. A common method
to control the path of continuous beams is to break the
translational invariance of the medium in the direction {\em
transverse} to the wave propagation direction, realizing a
graded-index (inhomogeneous) medium. The beam path is then
determined by the profile of the refractive index according to the
ray (eikonal) equation of Hamiltonian optics (see, for instance,
\cite{graded}). A similar method holds for discretized beams.
Graded-index waveguide lattices are usually realized by the
introduction of an inhomogeneous profile of the propagation
constants or of the coupling strengths for the various waveguides
\cite{Moison09,Fan09,Kartashov05}, and can find applications in
optical steering and focusing \cite{Moison09,Fan09}. For example,
plasmonic aperiodic waveguide arrays have been recently proposed to
realize deep sub-wavelength focusing and steering \cite{Fan09}.
However, the synthesis of a graded-index structure that yields a
desired beam path is a nontrivial issue. For beam propagation in
continuous media, the ray (eikonal) equation of Hamiltonian optics
shows that beam steering and refraction can be realized  by varying
the refractive index along the {\em longitudinal} propagation
direction, rather than in the transverse direction. This suggests
that beam path control for discretized light beams might be realized
in waveguide arrays {\em without breaking}  the periodicity in the
transverse direction. For example, steering of discrete optical
solitons in  optical lattices that fade away exponentially along the
propagation direction was predicted in Ref.\cite{Torner}. In this
Brief Report we propose a simple and rather flexible method to
control the path of a discretized beam in a homogeneous waveguide
lattice, which is based on longitudinal modulation of the coupling
constant. As compared to beam steering and refraction in
inhomogeneous graded-index arrays \cite{Fan09},
this approach enables to synthesize a rather arbitrary target path.\\
Before discussing the beam path control method for discretized
light, let us recall the refraction properties of paraxial beams in
a continuous medium with a refractive index $n$ which varies solely
along the paraxial direction $z$. For a monochromatic beam at
wavelength $\lambda$ (in vacuum), in the scalar approximation the
complex electric field amplitude $E(x,y,z)$ satisfies the Helmholtz
equation $\partial^2_z E+\nabla_t^2 E+k^2 n^2(z)E=0$, where
$\nabla_t^2$ is the transverse Laplacian and $k=2 \pi / \lambda$ the
wave number in vacuum. Assuming that $n(z)$ varies slowly over one
wavelength, by letting $E(x,y,z)=\psi(x,y,z) \exp \left[ ik \int_0^z
d \xi n(\xi) \right]$, in the paraxial approximation the envelope
$\psi$ satisfies the paraxial wave equation
\begin{equation}
i \partial_z \psi=-\frac{1}{2k n(z)} \nabla^2_t \psi
\end{equation}
with a $z$-dependence of the diffraction strength. The solution
$\psi(x,y,z)$ to Eq.(1), for an assigned initial field distribution
$\psi(x,y,0)$ at the $z=0$ plane,  can be simply obtained from the
corresponding solution of Eq.(1) in vacuum, in for $n=1$. In fact,
let $\phi(x,y,z)$ be the solution to Eq.(1) with $n=1$ and with
$\phi(x,y,0)=\psi(x,y,0)$. Then $\psi(x,y,z)=\phi \left(x,y,
\int_0^z d \xi / n(\xi) \right)$. Hence, since the center of mass of
the beam $\phi$ in vacuum propagates along a straight path defined
by the equations $x(z)=x(0)+\theta_x z$, $y(z)=y(0)+\theta_y z$,
where $\theta_x$ and $\theta_y$ are the paraxial beam angles at the
$z=0$ reference plane, the center of mass for the beam $\psi(x,y,z)$
propagates along a curved path defined by equations
$x(z)=x(0)+\theta_x \int_0^z d \xi / n(\xi)$, $y(z)=y(0)+\theta_y
\int_0^z d \xi / n(\xi)$. Thus a $z$-dependence of the diffraction
strength for continuous beams results into a non-straight beam
trajectory. Such a result can be extended to discretized light. As
the coupling rate $\kappa$ between adjacent waveguides in a
homogeneous array plays a similar role as the 'diffraction strength'
for continuous beams, a longitudinal change of $\kappa$ is expected
to curve the trajectory of the discretized beam. In fact, let us
consider a rather standard tight-binding model describing light
transport at wavelength $\lambda$ in a homogeneous waveguide array
with lattice period $a$ \cite{Christodoulides03,rev1} and with a
modulated ($z$-varying) coupling constant $\kappa(z)=\kappa_0 f(z)$,
\begin{equation}
i \dot{a}_n=-\kappa_0 f(z) (a_{n+1}+a_{n-1}),
\end{equation}
where $a_n$ is the modal amplitude of the light wave trapped in the
$n$-th waveguide of the array, $\kappa_0$ is a reference value of
the coupling constant, $f(z)$ is a modulation function, and the dot
stands for the derivative with respect to the longitudinal
propagation distance $z$. The solution $a_n(z)$ to Eq.(2), for an
assigned initial field distribution $a_n(0)$ at the $z=0$ input
plane,  can be simply obtained from the corresponding solution of
Eq.(2) for the non-modulated lattice, i.e. for $f(z)=1$. In fact,
let $\phi_n(z)$ be the solution to Eq.(2) with $f=1$ and with
$\phi_n(0)=a_n(0)$. Then $a_n(z)=\phi_n \left( \int_0^z d \xi f(\xi)
\right)$. Hence, since in a homogeneous array the center of mass of
the beam $\phi_n$ propagates along a straight path \cite{Longhi09},
then the center of mass for the discretized beam $a_n(z)$ in the
modulated lattice propagates along a curved path. Indeed, assuming
the normalization condition $\sum_n |a_n|^2=1$, after letting
$\langle n \rangle (z)=\sum_n n |a_n(z)|^2$ for the beam center of
mass, from Eq.(2) it readily follows that
\begin{equation}
\langle n \rangle (z)=\langle n \rangle (0) - 2 \kappa_0 \rho
\int_0^z d \xi f(\xi),
\end{equation}
where we have set $\rho \equiv {\rm Im} \left( \sum_n
a_n(0)^*a_{n-1}(0) \right)$. The parameter $\rho$ entering in Eq.(3)
is basically related to the tilting angle of the input beam. In
fact, for a broad input beam tilted at the angle $\theta$, one has
$a_n(0)=|a_n(0)| \exp(i \sigma n)$ and thus $\rho \simeq - \sin
\sigma$, where $\sigma= \pi \theta/ \theta_B$ and $\theta_B=
\lambda/(2a)$ is the Bragg angle. From Eq.(3) it follows that a
rather arbitrary path $\langle n\rangle (z)$ for the discretized
beam can be achieved by a suitable choice of the modulation function
$f(z)$, namely $f(z)=- (d \langle n \rangle /dz)/(2 \kappa_0 \rho)$.
It should be noted that nonlinear propagation of discretized
solitons in waveguide arrays with a longitudinally-modulated
coupling constant was previously considered in Ref.\cite{Lederer}
and shown to induce oscillations and decay of discrete solitons,
however the possibility to exploit the longitudinal modulation to
control the path of beams in the linear propagation regime was not
considered in such a previous work. Modulation of the coupling
constant can be effectively realized by either waveguide axis
bending or by out-of-phase modulation of the propagation constants
of adjacent waveguides (see, for instance,
\cite{Longhi06,DellaValle07,Longhi08OC,Kartashov09}). In such cases,
the coupled-mode equations for the modal amplitudes $c_n$ of light
trapped in the various waveguides read \cite{Longhi08OC}
\begin{equation}
i \dot{c}_n=-\kappa _0(c_{n+1}+c_{n-1})+q_n(z)c_n,
\end{equation}
where $\kappa_0$ is the coupling constant between adjacent
waveguides, and $q_n(z)=nG(z)$ for homogeneous arrays with axis
bending, or $q_n(z)=(-1)^n G(z)/2$ for straight arrays with
alternating modulation of the coupling constants. In the former
case, the modulation function $G(z)$ is related to the axis bending
profile $x_0(z)$ by  \cite{Longhi06}
\begin{equation}
G(z) = 2 \pi n_s \ddot{x}_0(z)a / \lambda,
\end{equation}
 where $n_s$ is the
substrate refractive index, whereas in the latter case $G(z)$
defines the alternating propagation constant mismatch between
adjacent waveguides. To establish an
 equivalence between the lattice models (2) and (4), let us assume a
 sinusoidal modulation function $G(z)$ with spatial period $\Lambda$ and slowly-varying
 amplitude $A$, i.e. $G(z)=A(z) \cos(2 \pi z / \Lambda)$, where $A$
 varies slowly over one spatial period $\Lambda$. After setting
 $c_n(z)=a_n(z) \exp[-i \int_0^z q_n(\xi)]$, Eq.(4) can be cast in
 the equivalent form
\begin{equation}
i \dot{a}_n=-\kappa_n (z) a_{n+1}-\kappa_{n-1}^*(z) a_{n-1},
\end{equation}
where we have set $\kappa_n(z)=\kappa_0 \exp \left[-i \int_0^z d \xi
G(\xi) \right]$ in case of waveguide axis bending, or
$\kappa_n(z)=\kappa_0 \exp \left[(-1)^n i \int_0^z d \xi G(\xi)
\right]$ in case of modulation of the propagation constants.
Assuming that the spatial modulation frequency $\Omega= 2 \pi /
\Lambda$ is larger than the coupling constant $\kappa_0$, at leading
order in a perturbative analysis of Eqs.(6) \cite{Longhi} the
evolution equations for the amplitudes $a_n$ take the form of
Eqs.(2) with a modulation function $f(z)$ given by
\begin{equation}
f(z)=J_0 \left( \frac{A(z)} {\Omega} \right)
\end{equation}
where $J_0$ is the Bessel function of first kind of zero order. In
particular, if the envelope $A(z)$ is varied such that $A/\Omega$
remains close to 2.405 (the first zero of Bessel $J_0$ function),
i.e. near the condition for suppression of evanescent tunneling
\cite{DellaValle07}, one has \cite{note}
\begin{equation}
f(z) \simeq -0.52 \left( \frac{A(z)}{\Omega}-2.405 \right).
\end{equation}
\begin{figure}
\includegraphics[scale=0.4]{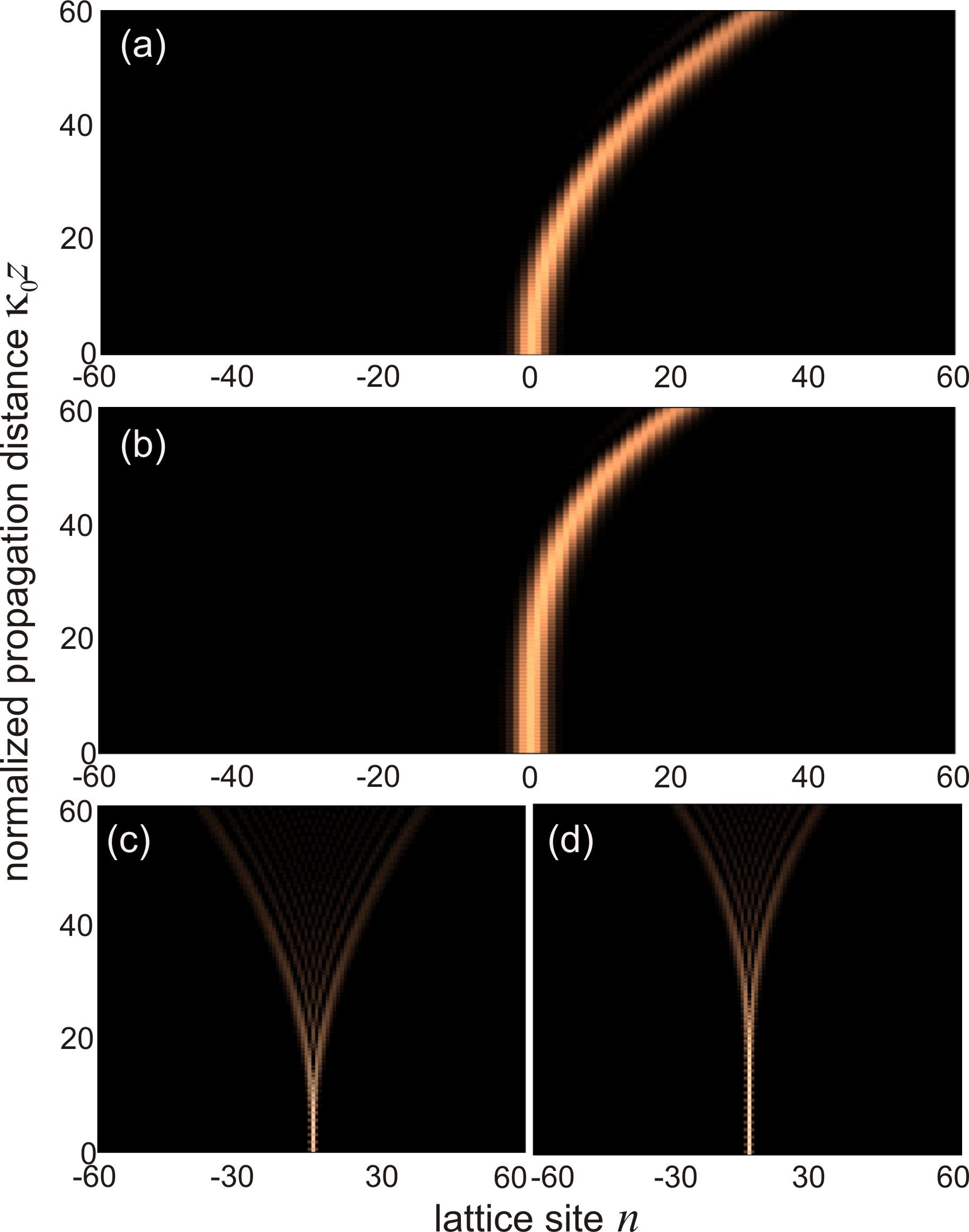}
\caption{ (color online)  Light evolution (snapshot of $|c_n|^2$
versus the normalized propagation distance $\kappa_0 z$) in a
homogeneous tight-binding lattice with a sinusoidally-bent  axis
with a normalized spatial frequency modulation $\Omega / \kappa_0 =
5$ and with amplitude $A(z)=\Omega (2.405-z/L)$ [in (a) and (c)],
and $A(z)=\Omega (2.405-z^2/L^2)$ [in (b) and (d)], where $L$ is the
array length. In (a) and (b) the array is excited by a tilted
Gaussian beam $c_n(0)=\exp(-n^2/9) \exp(i \pi n /2)$, whereas (c)
and (d) correspond to single waveguide excitation
$c_n(0)=\delta_{n,0}$. In (a) and (b) the beam follows a parabolic
and cubic path, respectively. In (c) and (d) light spreading is
faster than linear, corresponding to regimes of superdiffration.}
\end{figure}
This means that the modulation envelope $A(z)$ of axis bending or
propagation constant is just mapped into the modulation of the
coupling $f(z)$, which is in turn related to the beam trajectory via
the simple relation (3). Therefore, the synthesis of a quite
arbitrary beam path can be realized in a very simple way. For
instance, to realize a parabolic (cubic) path for a discretized
beam, according to Eqs.(3) and (8) a linear (parabolic) change of
$A(z)$ is required. This is shown, as an example, in Figs.1(a) and
1(b). In the figures, the evolution of the discretized intensity
distribution $|c_n(z)|^2$ along the array, as obtained by a
numerical analysis of Eqs.(4) for a tilted Gaussian input beam
distribution and assuming a periodically-bent axis $q_n(z)=n A(z)
\cos( \Omega z)$, is depicted for a linear [Fig.1(a)] and quadratic
[Fig.1(b)] variation of the amplitude $A(z)$, resulting in an
effective parabolic and cubic path of the beam, respectively. A
similar result is obtained by considering periodic out-of-phase
modulation of the propagation constants rather than periodic
waveguide axis bending.  It is interesting to notice that the
proposed scheme of beam path control also results in an effective
engineering of the discrete diffraction. For single waveguide
excitation at the input plane, the discrete diffraction pattern in a
homogeneous lattice with a constant coupling $\kappa_0$ evolves
according to $|a_n(z)|^2=J_n^2(2 \kappa_0 z)$, and thus light
spreads linearly with propagation distance following the same
ballistic transport law of electrons in tight-binding ordered
crystals \cite{balli}. In the waveguide array with the modulated
coupling constant, the spreading law can be engineered quite
arbitrarily. For example, in case of linear or parabolic change of
the amplitude $A(z)$, regimes of superdiffraction can be realized,
as shown in Figs.1(c) and (d). The equivalence between the lattice
models (2) and (4) has been established for a spatial modulation
frequency $\Omega$ larger than the coupling constant $\kappa_0$,
however it should be mentioned that beam steering can be achieved
even for slow modulation frequencies. For the case of axis bending
modulation, the expression of the beam path $\langle n \rangle (z)$
can be calculated in a closed form and reads $\langle n \rangle
(z)=\langle n \rangle (0)-2 \kappa_0 \rho \int_0^z d \xi
f(\xi)+s(z)$, where $f(z)$ is given by Eq.(7) and $s(z)$ is a
quasi-periodic function with period $\Lambda$ and with $s(0)=0$. As
for a fast modulation frequency $s(z) \rightarrow 0$ according to
the perturbative analysis, $s(z) $ is non-negligible when $\Omega$
is of the order or smaller than $\kappa_0$. However, since at the
planes $z=\Lambda, 2 \Lambda, 3 \Lambda,...$ $s(z)$ vanishes,
 a {\em coarse} beam steering control, at such discretized
planes, can be realized even for slow spatial modulation
frequencies. A {\em fine} beam path control requires, on the other hand, a fast modulation frequency.\\
To check the feasibility of the beam steering method, numerical
simulations of the full wave equation were performed. For the sake
of definiteness, the case of axis bending modulation was considered.
In the waveguide reference frame, the electric field envelope
$\psi(x,z)$ evolves according to the Schr\"{o}dinger-type wave
equation \cite{rev3,Longhi06}
\begin{equation}
i \lambdabar \partial_z \psi=-\frac{\lambdabar^2}{2n_s} \partial_x^2
\psi+V(x) \psi+n_s \ddot{x}_0(z) x \psi,
\end{equation}
\begin{figure}
\includegraphics[scale=0.4]{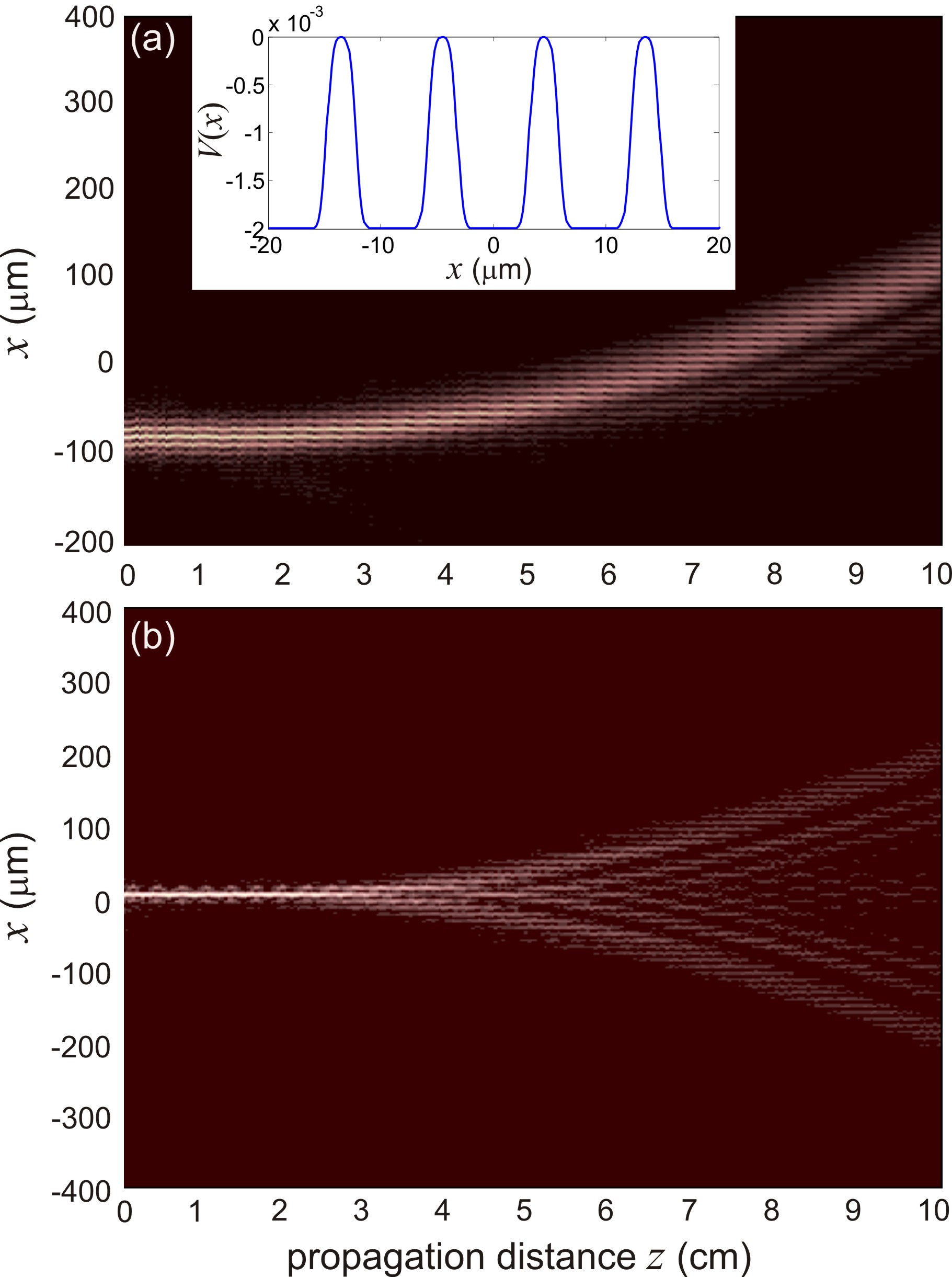}
\caption{ (color online) (a) Parabolic path of a broad Gaussian beam
in a sinusoidally-curved homogeneous waveguide array with
linearly-varying modulation amplitude. The figure shows the
evolution of the light intensity $|\psi(x,z)|^2$ versus propagation
distance $z$ for Gaussian input beam excitation (tilting angle
$\theta=\theta_B/2$). The inset shows the behavior of the optical
potential $V(x)=n_s-n(x)$ of the one-dimensional lattice. The values
of other parameters are given in the text. (b) Same as (a), but for
single waveguide excitation of the array. Note that light spreads
quadratically versus propagation distance (superdiffractive regime),
rather than linearly as in a homogeneous array with constant
coupling.}
\end{figure}
 where $\lambdabar=\lambda/(2 \pi)$ is the reduced wavelength of injected light, $V(x) \simeq n_s-n(x)$ is the periodic lattice
potential, $n(x)$ is the refractive index profile of the array, and
$x_0(z)$ is the profile of axis bending. Equation (9) was integrated
by a standard pseudospectral split-step method for a typical
waveguide lattice manufactured by femtosecond laser writing and
excited at $\lambda=633$ nm \cite{Szameit09,Kartashov09}. The
refractive index profile of the array used in numerical simulations
is shown in the inset of Fig.2(a). The corresponding band diagram,
computed by a standard plane-wave expansion method, shows that the
lowest band is well separated from the higher-order bands and its
dispersion curve is well fitted by a sinusoidal curve. Hence,
provided that the array is excited by an input beams tilted at an
angle smaller than the Bragg angle and for a spatial modulation
frequency small enough to avoid coupling to higher-order bands
(radiation losses), the beam evolution in the lattice  turns out be
well described by the tight-binding lattice model (4). In the
example shown in Fig.2, the axis bending profile $x_0(z)$ has been
chosen to realize a parabolic beam path as in Fig.1(a), namely we
assumed $x_0(z)=B(1-z/2L) \cos (2 \pi z / \Lambda)$, where $L$ is
the length of the waveguide array and $B=2.405 \times \Lambda
\lambda/(4 \pi^2 n_s a)$ is the bending amplitude corresponding to
tunneling inhibition (dynamic localization \cite{Longhi06}). The
spatial period $\Lambda$ of the modulation used in the simulations
is $\Lambda=3 \; {\rm mm}$, corresponding to $B \simeq 9 \; \mu$m
for a bulk refractive index $n_s=1.42$ and lattice period $a= 9 \;
\mu$m. Figure 2(a) shows the evolution of light intensity
$|\psi(x,z)|^2$ along the array for a Gaussian-shaped input beam
with spot size $w= 21.6 \; \mu$m, tilted at half of the Bragg angle
($\theta=\theta_B/2 \simeq 1^{\rm o}$), i.e.
$\psi(x,0)=\exp(-x^2/w^2) \exp(i \pi x / 2a)$. According to the
tight-binding model, the parabolic path followed by the beam is
clearly visible, with negligible radiation losses induced by axis
bending. Figure 2(b) shows the evolution of light intensity when a
single waveguide of the array is excited in its fundamental mode at
the input plane, leading to a
superdiffraction regime for light spreading as in Fig.1(c).\\
To conclude, a flexible and simple method for the control of the
path of discretized light beams in homogeneous waveguide arrays,
based on longitudinal modulation of the coupling constant, has been
theoretically proposed. As compared to beam steering and refraction
control achievable in graded-index waveguide arrays, the proposed
method enables to synthesize rather arbitrary target paths, and
could be therefore of potential interest for beam steering
applications in discrete photonics. Owing to the quantum-optical
analogy between light transport in waveguide arrays and coherent
electronic or matter wave transport in solid-state or matter wave
systems \cite{rev3}, the proposed method could be of interest beyond
discrete optics. For example, it could be applied to control the
path of coherent electronic wave packets as well as to realize
superdiffusive coherent electronic transport in ac-driven quantum
dot arrays \cite{Hanggi}.

 This work was supported by the Italian MIUR (Grant
No. PRIN-20082YCAAK).


\begin{thebibliography}{}
\bibitem{note}
Linearization (8) of the Bessel $J_0$ function around its first zero
$A/\Omega=2.405$ is a very well approximation for $A/\Omega$ varying
in the whole range $(1,2.6)$.
\end{thebibliography}
\end{document}